\documentclass[]{raa}            % referee version: for submission

\usepackage{graphicx,times}             %for PS/EPS graphics inclusion, new
\usepackage{natbib}
\usepackage{multirow}
\usepackage{amssymb,amsmath}
\usepackage{multirow}
\bibpunct{(}{)}{;}{a}{}{,}
\usepackage{mathrsfs}

\usepackage[pagebackref=true]{hyperref}

\begin{document}

  \title{Understanding the predication mechanism of deep learning through error propagation among parameters in strong lensing case
}
%   \subtitle{I. Place Your Subtitle Here}

 %  \volnopage{Vol.0 (20xx) No.0, 000--000}      %%preserved for Editor. DOn't remove!
 %  \setcounter{page}{1}          %%starting page, preserved for Editor. DOn't remove!

   \author{Xilong Fan %(周爱英) %% Put your Chinese name in "( )" if you like. Note to open line 11 "\usepackage[UTF8]{ctex}"
      \inst{1}
   \and Peizheng Wang
      \inst{2}
   \and Jin Li
      \inst{3}
    \and Nan Yang
      \inst{4}
   }

   \institute{School of Physics and Technology, Wuhan University, Wuhan, Hubei 430072, China\\
        \and
             School of Software Technology, Zhejiang University, Ningbo 315048, China\\
        \and
             Department of Physics, Chongqing University, Chongqing 401331, China,  \href{mailto:cqujinli1983@cqu.edu.cn}{cqujinli1983@cqu.edu.cn}\\
        \and
            Department of Electronical Information Science and Technology, Xingtai University, Xingtai 054001, China, \href{mailto:cqunanyang@hotmail.com}{cqunanyang@hotmail.com}\\
\vs\no
   {\small}}

\abstract{The error propagation among estimated parameters reflects the correlation among the parameters. We study the capability of machine learning of "learning" the correlation of estimated parameters. We show that machine learning can recover the relation between the uncertainties of different parameters, especially, as predicted by the error propagation formula. Gravitational lensing can be used to probe both astrophysics and cosmology. As a practical application, we show that the machine learning is able to intelligently find the error propagation among the gravitational lens parameters (effective lens mass $M_{L}$ and Einstein radius $\theta_{E}$) in accordance with the theoretical formula for the singular isothermal ellipse (SIE) lens model. The relation of  errors of lens mass and Einstein radius, (e.g. the ratio of standard deviations $\mathcal{F}=\sigma_{\hat{ M_{L}}}/ \sigma_{\hat{\theta_{E}}}$) predicted by the deep convolution neural network are consistent with the error propagation formula of SIE lens model. As a proof-of-principle test, a toy model of linear relation with Gaussian noise is presented. We found that the predictions obtained by machine learning indeed indicate the information about the law of error propagation and the distribution of noise. Error propagation plays a crucial role in identifying the physical relation among parameters, rather than a coincidence relation, therefore we anticipate our case study on the error propagation of machine learning predictions could  extend to other physical systems on  searching the correlation among parameters.
}

   \authorrunning{Xilong Fan, et. al}            %author_head in even pages
   \titlerunning{Error propagation in strong lensing case}  % title_head in odd pages

  \maketitle
%% The author head (on even pages) and the title head (on odd pages) will be
%% automatically extracted from \author{} and \title{}. Whenever the title is too long,
%% you will be asked to supply a shorter one by inserting either \authorrunning{} or
%% \titlerunning{} before \maketitle. Anyway, you can specify your own heads.
%%
%%
%% Note: In the following text body of your manuscript, please note several differences from
%%       other major journals:
%% (1) \subsection{Please Capitalize the First Letter of Each Notional Word in Subsection Title}
%% (2) Please Capitalize the First Letter of Each Notional Word in all tables' captions

%
%________________________________________________ sections below
%
\section{Introduction}           %% first-level sections will be auto-capitalized
\label{sect:intro}

Since 1979 \citep{DiscoverLens}, gravitational lensing effects have been used as a practical approach in numerous researches of astrophysics and cosmology (e.g. see recent reviews \cite{2019RPPh...82l6901O,2022ChPhL..39k9801L}). As an important role in astronomical research, gravitational lens can generate the multiple images of galaxies, quasars and supernovae, Einstein Cross and Einstein ring and so on, which contain very important information about luminous objects \citep{galx1,galx2,galx3,galx4,galx5,quas1,quas2}. Furthermore, gravitational lensing effects also play an important role in the study of cosmology. By using the gravitational lens, astronomers and cosmologists can determine the distribution of baryonic matter and dark matter in galaxies and clusters of galaxies more precisely, and then determine some important parameters of cosmology \citep{dark1,dark2,cosmy1,lensGW1}. 

Although many of the lensing systems have been found through the traditional searches (e.g., \cite{opLenssearch}), with the rapidly increasing data sets, the enhancement of automated methods to discover lens candidates and estimate the relationship among the parameters become highly necessary \citep{nature}. 
Besides searching candidate, modeling is executed by running maximum likelihood algorithms that were computationally expensive (e.g., \cite{ConventionalModel1,ConventionalModel2,ConventionalModel3}), and the traditional parameter estimation methods are time consuming \citep{tradlens}. Convolutional neural networks (CNNs), known as a class of deep learning networks, can be trained to identify characteristics of specific images.  Recently, CNNs has been used to study lens modelling as a more efficient parametric methods \citep{nature,recommend2,natureauthor2}. Furthermore, the authors of \cite{nature} have extended the work to estimate the uncertainties in parameters with neural networks \citep{BayesianPE1}, which was produced by using dropout techniques that evaluates the 
deep neural network from a Bayesian perspective \citep{BayesianPE2,recommend1}.
Some latest related work \citep{2021ApJ...909..187W,2021ApJ...910...39P} have demonstrated Neural Networks can be used as a powerful tool for uncertainties inference.  The main purpose of their work is to improve the prediction accuracy: by eliminating some unrepresentative prior deviations of the training set, the deviation of the predicted results of the testing dataset will not be affected by the deviations.   %\textcolor{red}{In Glasgow ml paper\cite{??}, the network learns the posterior samples to predicate ... }
 
Whether the uncertainties of estimated parameter by the Network reflects the a  correlation among the parameters plays a important role in  understanding the predication  mechanism.
Being different from the above works on estimation error, in this paper we focus on  the relation  of errors  of  prediction results by  machine learning. Namely, we test whether  the error relation can reflect the correlation among the predicted objects through the prediction errors from deep neural networks (DNNs). We compare the estimation results of the effective lens mass (which can not be observed directly from the lens image) and Einstein radius (which can  be approximately measured  from the lens image) to find the error propagation among the parameters prediction of machine learning, then to find the potential relationship between the two parameters. To our knowledge, the current work is the first one that shows that machine learning can recover the relation between the uncertainties of different parameters, especially, as predicted by the error propagation formula. 
In fact, the correlation of the uncertainty in each parameter can be learned by general Neural Networks automatically, which will be demonstrated by means of machine learning on a linear relation toy model and strong lens data in Section \ref{restuls}. The detailed information about simulation and results in two models (toy model and lens model) is discussed in Section \ref{Simu-res}. Summary is drawn in Section \ref{summary} with some additional discussion.

%% Authors can give a citation as 'Michel et al. 1992'.
%% You may also use \cite, \citep and \citet for citation, and use Table~1 or Figure~1
%% and so forth. Using \ref and \label for cross-references of Tables/Figures
%% is a good way in adjusting/adding/removing text, tables or figures.

\section{Error propagation among  parameters}\label{restuls}
\begin{figure} %[!htb]
	\centering
	\includegraphics[height=10cm,width=16.0cm]{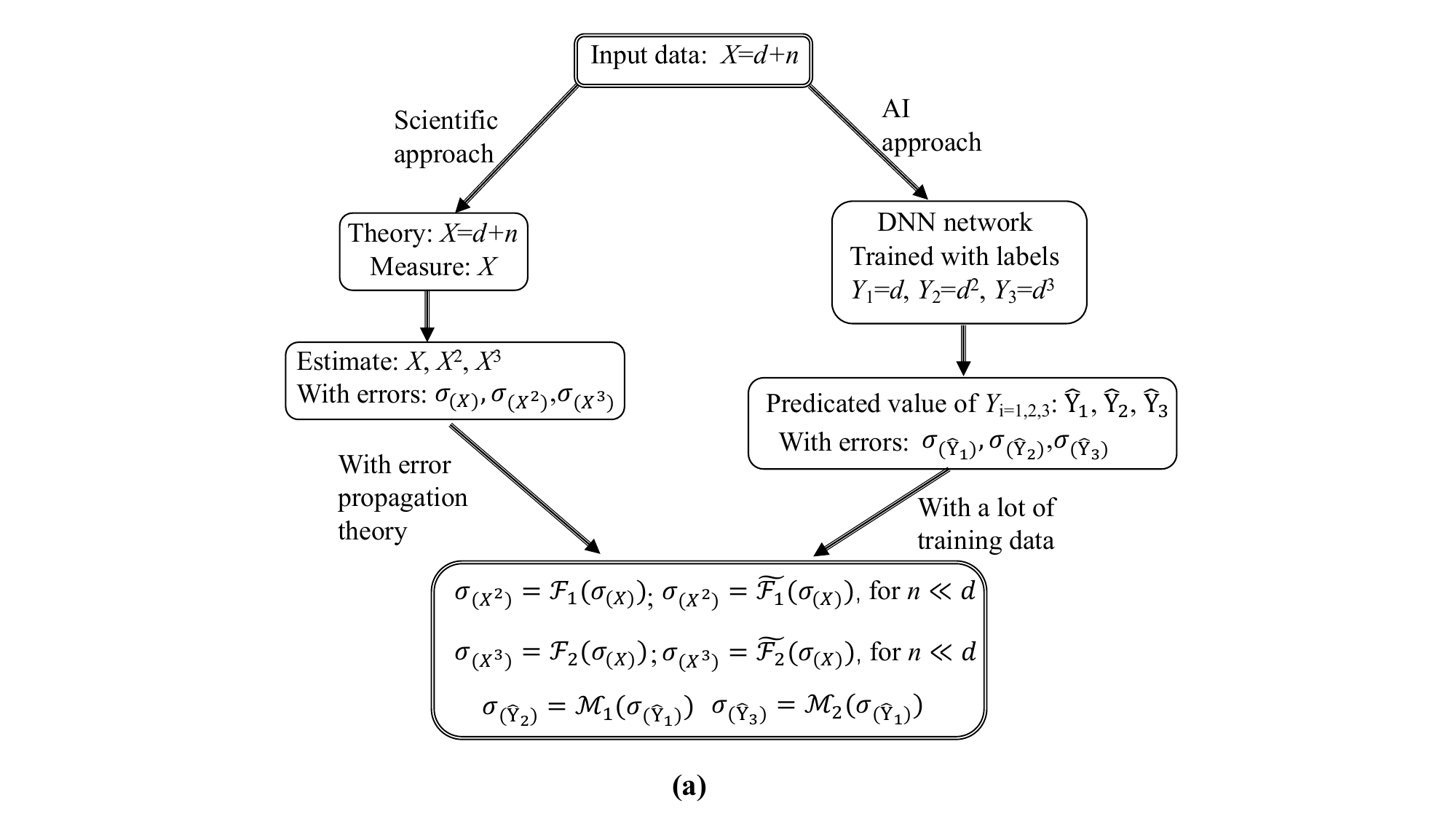}
		\includegraphics[height=12cm,width=16.0cm]{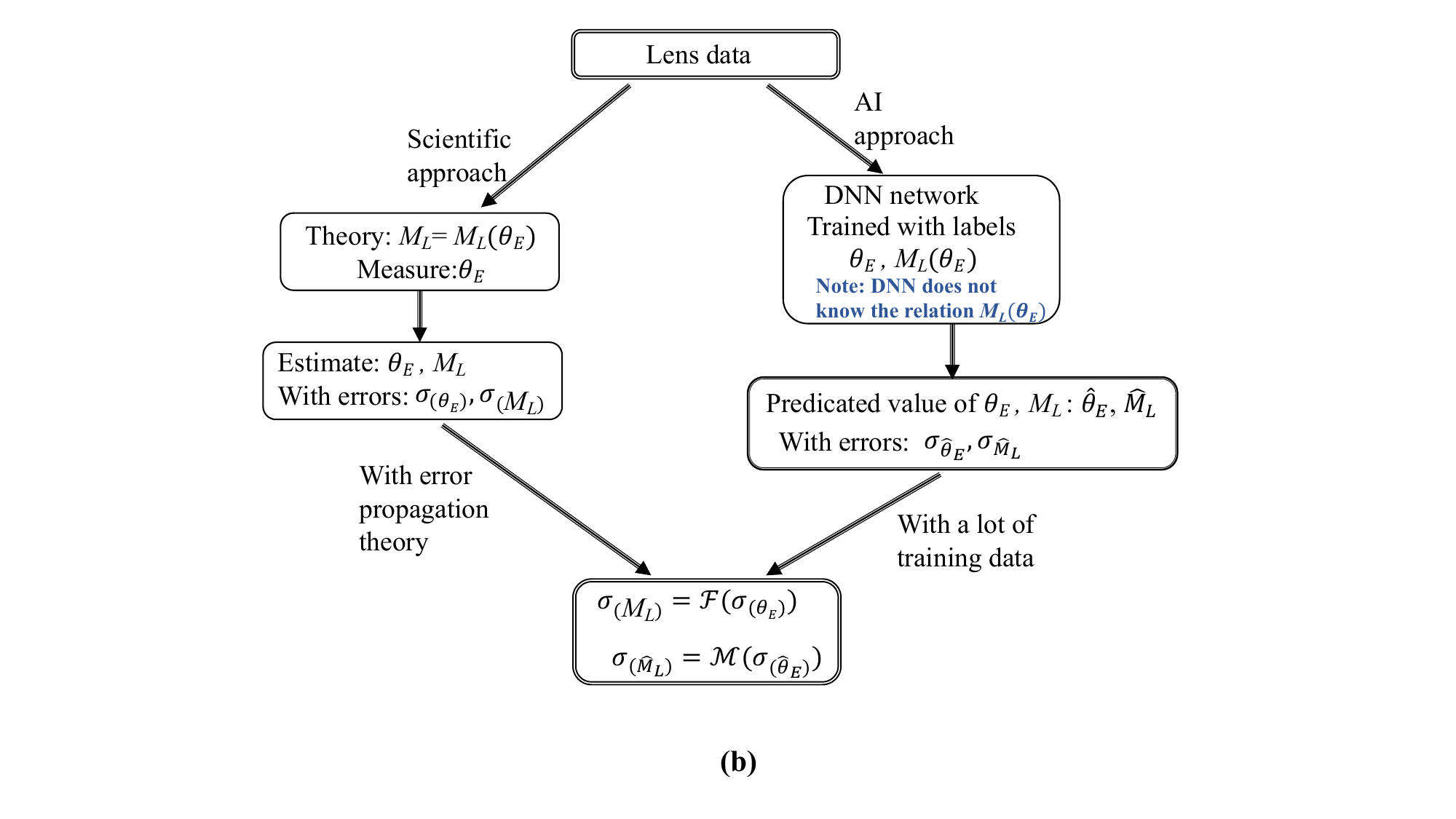}
	\caption{Schematic diagram  of two approaches for toy model(a) and lens model(b).  In AI approach,  labels for the lens case are produced by a relation,  but the relation itself are not presented to the network in training progress. We found $\mathcal{M}_1 (\sigma_{(\hat{Y_2})}, \sigma_{(\hat{Y_1})})\sim\mathcal{F}_1$  and $\mathcal{M}_2 (\sigma_{(\hat{Y_2})}, \sigma_{(\hat{Y_1})})\sim\mathcal{F}_2$ for the toy model, and $\mathcal{M} \sim \mathcal{F}$ for the lens model (see detail results  in Section \ref{restuls}). }
	\label{explian}
\end{figure}

Traditionally, given the known physical relation of parameters or an analytic likelihood function, the relation of uncertainties of parameter estimation is presented by the error propagation formula, the  Fisher matrix approach, or a Bayesian posterior distribution of multi-parameters. The error propagation formula is quite common and most simple approach when one can not directly measure some parameters. By using a particular relation of two parameters, the differentiation law and the Taylor expansion, one can derive the error propagation formula of two parameters on their standard deviation $\sigma$:  
 \begin{equation}
\sigma^2(y)= \frac{\partial y}{\partial  x}  \, \sigma^2(x).
\end{equation}
The effective lens mass $M_{L}$ (see definition in Eq.(\ref{m_e})) in a strong lens system is an example. In traditional estimation approach, people could directly measure the $\theta_{E}$ and estimate the lens mass $M_{L}$ based on a lens model. The errors of $M_{L}$ (e.g. the standard deviation $\sigma_{M_{L}}$) is calculated through the error propagation law with the error of $\theta_{E}$ ($\sigma_{\theta_{E}}$). 
While, the Neural Networks do not need this known relation to get $\frac{\partial y}{\partial x}$, since in the supervised-training step one can directly design any label. Then the Neural Networks could directly show the results of $\sigma^2(y)$ and $\sigma^2(x)$. Note that, when individual label relates to individual parameter, the relation of parameters is not indicated in the training process. We highlight the difference of the error propagation  approach and the Neural Networks  approach in Fig \ref{explian}.

We will demonstrate the error propagation of Neural Networks in two cases: a linear relation toy model with Multi-layer Perceptron based networks in Section \ref{Toymodel}  and the lens model with convolution-based networks in Section \ref{sec:DandA}.  Here we summary the main results using the symbols in Table \ref{table1}. 
Taking the lens model into consideration, the relation of $\sigma^A_{\hat{\theta_{E}}}$ and $\sigma^B_{\hat{M_{L}}}$ from two Networks do not have clear relation, science the Network is not trained with information on the data noise distribution and we do not know the systematical error of the Network itself. So it is not trivial to check if the relation of $\sigma^C_{\hat{\theta_{E}}}$-$\sigma^C_{\hat{M_{L}}}$ (or $\sigma^C_ {\hat{Y_1}}$-$\sigma^C_{\hat{Y_2}}$) derived from a Network following the error propagation law. In fact, if the predication error is only from the Network itself (e.g. the label value is exactly equal to the data value), the $\sigma^C_{\theta_{\hat{E}}}$-$\sigma^C_{\hat{M_{L}}}$ relation does not follow the error propagation law assuming Gaussian noise.  On the other hand, if the data noise is dominated, the $\sigma^C_{\theta_{\hat{E}}}$-$\sigma^C_{\hat{M_{L}}}$ relation follows the error propagation law (see detail in Fig \ref{figerror}). This consistency is a puzzle for us, since one did not label the $\theta_{E}$-$M_{L}$ relation in the training process but just separately label the true parameters of $\theta_{E}$ and $M_{L}$.

% \usepackage{multirow}
\begin{table}[]
\centering
\caption{Symbol description. Error relations from Networks are compared with the error propagation formulas. Network A and B are designed for one parameter in error propagation formulas,while Network C is for all parameter in error propagation formulas. All the errors are listed in the corresponding columns. In the network approach neither information on  noise nor the relation form of parameters are provided to Networks. For the toy model, predication errors of Network C (Eq.(\ref{eq_error3})) in the no-noise data ($n=0$) case does not follow the error propagation formula at low noise limit (Eq.(\ref{eq_error2})), while for the noise data ($n\ne 0$) cases, relations of errors from Network A, B and C are all consistent with the error propagation formula (Eq.(\ref{eq_error1})). 
	%\textcolor{red}{ In higher Gaussian noise level, the Network C {\bf is better than(这个结论的得出是针对公式3, 但是其实和公式2完全符合, 没有“更好”这个结论？)} the error propagation law on revealing the relation of $\sigma^C_{Y_1}$-$\sigma^C_{Y_2}$, $\sigma^C_{Y_1}$-$\sigma^C_{Y_3}$ which indicates that the Network could also ``learn" the noise properly.\textcolor{green}{应该是在高的噪声水平下, 模型拟合导致的误差相较于噪声导致的误差可以忽略不计, 所以在图中看起来高噪声下要好。} }
 For lens model, only the relation $\sigma^C_{\hat{\theta_{E}}}$-$\sigma^C_{\hat{ M_{L}}}$ by Network C is consistent with the predication of the error propagation formula based on the SIE lens model (Eq. (\ref{err_p2})). Symbols with hat represent the results from Networks.}
\label{table1}
\begin{tabular}{c|cc|c|c|c|c}
\hline
Models   & \multicolumn{2}{c|}{Parameters} & Network A  &  Network B  & Network C  & Error propagation consistency\\ \hline
\multirow{2}{*}{lens}  &   Parameter I  &	$\theta_{E}$  &   $\sigma^A_{\hat{\theta_{E}}}$   & -------  &$\sigma^C_{\hat{ \theta_{E}}}$  & \multirow{2}{*}{C} \\
 &    Parameter II   & $M_{L}\propto \theta^2_{E}$&    -------  &$\sigma^B_{\hat{ M_{L}}}$ & $\sigma^C_{\hat{ M_{L}}}$   & \\ \hline   \hline
\multirow{3}{*}{Toy} &  Parameter I& $Y_1 = d $& $\sigma^A_{\hat{ Y_1}}$ &-------& $\sigma^C_{\hat{ Y_1}}$ &  \\
&   Parameter II$_1$& $Y_2 = d^2 $&------- &$\sigma^B_{\hat{ Y_2}}$ & $\sigma^C_{\hat{ Y_2}}$  & \multirow{2}{*}{A,B,C} \\
 &  Parameter II$_2$& $Y_3 = d^3 $&------- &$\sigma^B_{\hat{ Y_2}}$ & $\sigma^C_{\hat{ Y_3}}$  &\\ \hline
\end{tabular}
\end{table}

\subsection{Toy Model}\label{Toymodel}
We design a linear relation with  Gaussian noise $n$ as follows:
 \begin{equation}\label{eq_dataxy}
\left\{ \begin{array} { l } 
X=d+n\\
Y_i=d^i ; \quad  i=1,2,3 \\ 
\hat{Y_i}=Y_i+N_i\\
 \end{array}  \right. 
 \end{equation}
 where $d$ is a arbitrary value and $n$ is a Gaussian noise $n\sim N(0,\sigma)$, $X$ is a input data with three labels $Y_i$, $\hat{Y}_i$  is the predicted value of $Y_i$, and $N_i$ is the prediction error from the neural network. For each $X$, we assign a label set $\{Y_1, Y_2, Y_3\}$, which is in values $\{d, d^2, d^3\}$. By this type labels, we not only try to check if the Network could overcome the Gaussian noise $n$ and predict the true value $\{d, d^2, d^3\}$ from $X$, but also to investigate the relation of errors of predication $\{\sigma_{(\hat{Y_1})},\sigma_{(\hat{Y_2})},\sigma_{(\hat{Y_3})}\}$. The predicted value $\{\hat{Y_1},\hat{Y_2},\hat{Y_3}\}$ given by the neural network can be regarded as a function of the testing variable $X$, which is determined by an un-known predication mechanism of the neural network. Therefore the properties of predication errors $\{N_1,N_2, N_3\}$, e.g. the distribution of them, are unclear.   
 
However, we could directly compare the predication results of $\{\sigma_{(\hat{Y_1})}, \sigma_{(\hat{Y_2})}, \sigma_{(\hat{Y_3})}\}$ (labeled as a function $\mathcal{M}$ among them in below) with the well-defined error propagation relations of $\{X, X^2, X^3\}$ (labeled as a function $\mathcal{F}$ in below).

For the toy model, we can get the relation of errors of $\{X, X^2, X^3\}$ through the definition of standard deviation by the error propagation law:
%$$
%E_1 \equiv \frac{\sigma_{(X^2)}}{\sigma_{X}}=\frac{\sigma_{[(\textcolor{red}{x}+n)^2-\textcolor{red}{x}^2]}}{\sigma_{n}}=\frac{\sigma_{(2\textcolor{red}{x}%n+n^2)}}{\sigma_{n}}=\frac{\sqrt{\sigma_{(2xn)}^2+\sigma_{(n^2)}^2+2\text{Cov}(2xn,n^2) }}{\sigma_{n}}
%$$
 \begin{equation}\label{eq_error1}
\left\{ \begin{array} { l } 
 \frac{\sigma_{(X^2)}}{\sigma_{(X)}}=\frac{\sqrt{4d^2\sigma_{(n)}^2+\sigma_{(n^2)}^2 }}{\sigma_{(n)}}=\sqrt{4d^2+2\sigma^2} \equiv \mathcal{F}_1  \\
\\ 
\frac{\sigma_{(X^3)}}{\sigma_{(X)}}=\sqrt{9d^4+36d^2\sigma^2+15\sigma^4} \equiv \mathcal{F}_2 \end{array} \right. 
\end{equation}
where $\sigma_{(n)}=\sigma$ is the standard deviation of the Gaussian noise $n$ and $\sigma_{(n^2)}=2 (\sigma_{(n)})^4 = 2 \sigma^4 $, which allows us to use the co-variance ($\text{Cov}$) properties of any variable with zero mean,
$$
 \begin{array} { l } 
\text{Cov}(2 d n,n^2)=2 \, d \, \text{Cov}(n,n^2)=0
\end{array}.
$$
At low noise limit e.g. $\sigma << d $, we have:
 \begin{equation}\label{eq_error2}
\left\{ \begin{array} { l } 
 \frac{\sigma_{(X^2)}}{\sigma_{(X)}} \sim 2|d|   \equiv \widetilde{\mathcal{F}}_1  \\
\\ 
  \frac{\sigma_{(X^3)}}{\sigma_{(X)}} \sim 3d^2   \equiv  \widetilde{\mathcal{F}}_2
 \end{array} \right. 
\end{equation}
It is worth to check if the noise $n$ could affect the relation of errors ${\sigma_{(\hat{Y_1})},\sigma_{(\hat{Y_2})},\sigma_{(\hat{Y_3})}}$ from the networks:
\begin{equation}\label{eq_error3}
\left\{ \begin{array} { l } 
 \frac{\sigma_{(\hat{Y_2})}}{\sigma_{(\hat{Y_1})}} \equiv \mathcal{M}_1 (\sigma_{(\hat{Y_2})}, \sigma_{(\hat{Y_1})})  \\
\\ 
 \frac{\sigma_{(\hat{Y_3})}}{\sigma_{(\hat{Y_1})}}  \equiv \mathcal{M}_2 (\sigma_{(\hat{Y_3})}, \sigma_{(\hat{Y_1})}) 
 \end{array} \right. 
\end{equation}

\subsection{The lens model} \label{sec:DandA}
For the lens system $X=d+n$, the data $X$ is an observed image (e.g. see the first supernova  lens image in  \cite{2015Sci...347.1123K}). The image could be reconstructed by a lens model $d$, while the noise $n$ is more complex than a Gaussian noise. We adopt a SIE lens model, which is described by five parameters: 
the values of Einstein radius $\theta_{E}$, the complex ellipticity ($\epsilon_{x}, \epsilon_{y}$) and the position of lens centre ($x, y$). For both training and testing data, those parameters are drawn from the uniform distribution shown in Table \ref{Errors} with different random seeds.
The network could directly predicate those five parameters \{$\theta_{E}$, $\epsilon_{x}$, $\epsilon_{y}$, $x$, $y$\}. 
For the lens model, the effective lens mass $ M_{L}$  (the mass enclosed inside the
Einstein radius) is related to the Einstein radius $\theta_{E}$ (\cite{1992grle.book.....S}):
\begin{equation}\label{m_e}
M_{L} =\frac{c^{2}D}{4G }\theta_{E}^{2} =\frac{c^{2}D_{l}D_{s}}{4G D_{ls}} \theta_{E}^{2},
\end{equation}
where $D_{l}, D_{s}$ are the angular diameter distance of lens and source respectively, $D_{ls}$ is the angular diameter distance between lens and source. Traditionally, those distances is inferred by redshifts of lens and source through a cosmology model (see a recent review in \cite{2019RPPh...82l6901O}). Therefore, the model for the Network could also be described by parameters \{$M_{L}$, $\epsilon_{x}$, $\epsilon_{y}$, $x$, $y$\}, if we can measure the redshift of the source and lens by emission lines. 

Unlike Einstein radius, lens mass is strongly model dependent parameter and could not be observed by telescope directly.  However for machine learning approach, all the parameters could be predicted from the input directly.
Shown in the toy model case, deep learning is able to extract deep information from input and predict any designated source parameters. We attempt to check whether it can also learn the association among parameters indicated in the propagation of uncertainty of predictions for the lens model. Here we take $\theta_{E}$ and $M_{L}$ into concern (Eq. \ref{m_e}), and the predicted errors ratio by the theoretical error propagation formula is:
\begin{equation}\label{err_p}
\mathcal{F} \equiv \frac{\sigma_{  M_{L} }}{\sigma_{ \theta_{E}}}=c\sqrt{\frac{D}{G}}\sqrt{M_{L}}.
\end{equation}
Since we only try to recover the relation of errors of two parameters ($M_{L}$ and $\theta_E$),
here we do not infer the redshifts of lens and source, but just assuming $D_{l}$, $D_{s}$  and $D_{ls}$ are  known constant as did in \cite{nature}, e.g. fixed $z_s=0.5$ and $z_l=0.2$ (see more detail in Section \ref{LensPre}). The parameters \{$\epsilon_{x}$, $\epsilon_{y}$, $x$, $y$\} are not fixed for the training  and  testing processes to make sure that all results could be used as a astrophysical application in our next work. It should be also noted that: although we generate the labels of $ M_{L}$ and $\theta_{E}$ for the training data sets with the SIE lens model, the prediction processes of DNN are not informed of any relationship of these parameters.   

Similarly to error relation of $\hat{Y_1}$ and $\hat{Y_2}$ in the toy model, the error relation of $\hat{M_{L}}$ and $\hat{\theta_{E}}$ is the target of this section. 
Therefore we could also design three label sets: \{$\theta_{E}$, $\epsilon_{x}$, $\epsilon_{y}$, $x$, $y$\}, \{$ M_{L}$, $\epsilon_{x}$, $\epsilon_{y}$, $x$, $y$\}, and \{$ M_{L}$, $\theta_{E}$, $\epsilon_{x}$, $\epsilon_{y}$, $x$, $y$\}, for one observed image $X$.  
For the same data $X$, three networks shown in Table \ref{Errors} are adopted to predicate the common  parameters  \{$\epsilon_{x}$, $\epsilon_{y}$, $x$, $y$\}, and 
(i) Network VGG16($\theta_{E}$) for \{$\theta_{E}$, $\epsilon_{x}$, $\epsilon_{y}$, $x$, $y$\} also gives the prediction $\hat{\theta_{E}^A}$ with the predication error $N^{A}_{\theta_{E}}$;
(ii) Network VGG16($M_{L}$) for \{$ M_{L}$, $\epsilon_{x}$, $\epsilon_{y}$, $x$, $y$\}  also gives the prediction $\hat{M_{L}^{B}}$ with the predication error $N^{B}_{M_{L}}$;
(iii) Network VGG16($\theta_{E}, M_{L}$) for \{$ M_{L}$, $\theta_{E}$, $\epsilon_{x}$, $\epsilon_{y}$, $x$, $y$\} also gives the prediction $\hat{\theta_{E}^C}$ and $\hat{M_{L}^{C}}$ with the predication errors $N^{C}_{\theta_{E}}$ and $N^{C}_{M_{L}}$.
Predication errors $N_i^j$ (i$=\theta_{E}$, $M_{L}$; j$=$A, B, C)
are caused by lens data noise and unknown network prediction mechanism, therefore we do not know the statistical properties of $N^{C}_{M_{L}}$ and $N^{C}_{\theta_{E}}$. 
 
\subsection{Neural Networks}

{\bf{For toy model:}} We train the networks and predict $\{\hat{Y_1}, \hat{Y_2}, \hat{Y_3}\}$ with a two-layer fully connected neural network 
one layer has 32 fully-connected units and another has  1  fully-connected units for Network A and B while 3 fully-connected units for Network C. We choose mean squared error (MSE) and Rectified Linear Unit (ReLU) as the loss function and activation function respectively, optimize the network by ADAM algorithm, set the batch size to be 512, and adjust the learning rate to be $10^{-3}$ for the $100$ epochs. Shown in Table \ref{table1}, we  design three types of networks for different predication parameter sets: Network $A$ for only $\{\hat{Y_1}\}$, Network $B_1$ for $\{\hat{Y_2} \}$, Network $B_2$ for $\{\hat{Y_3} \}$, Network $C$ for $\{\hat{Y_1},\hat{Y_2},\hat{Y_3}\}$.

\begin{figure*}[htbp]
	\centering
	
	\includegraphics[height=6.5cm,width=16.0cm]{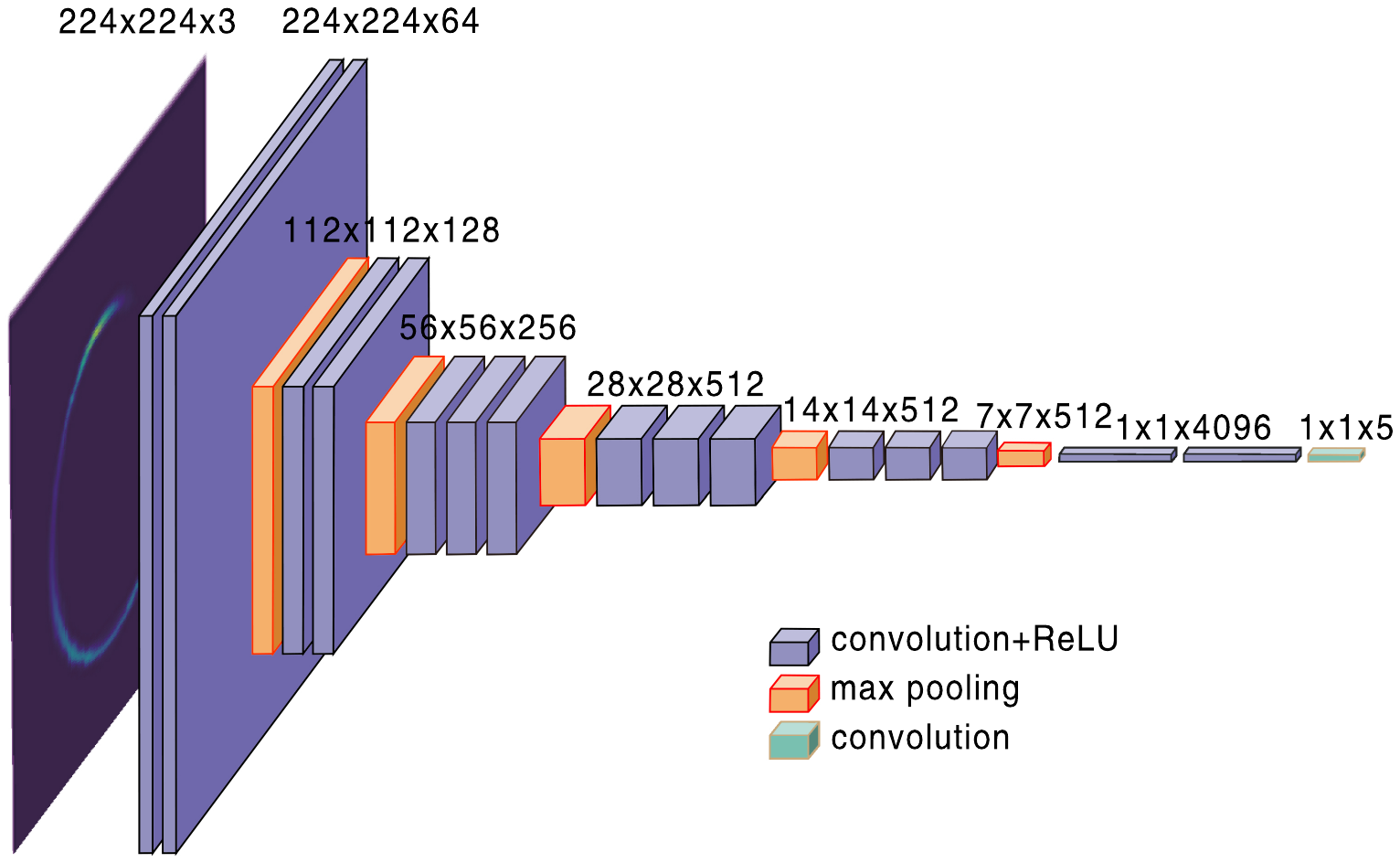}
		
	\caption{The structure of the VGG16 used in this work.}
	\label{VGG16}
\end{figure*}

{\bf{For lens model:}} Besides the AlexNet used in \cite{nature}, we adopt the VGG16 network \citep{VGG} to predict the lens parameters. VGG16 network is a common deep learning structure and sometimes outperforms AlexNet on computer vision tasks. We adjust the final layer to the fully connected layer to regress the parameters in VGG16. The structure of our VGG16 network is shown in Fig.\ref{VGG16}.

In training process, we choose averaged mean squared error (MSE) and Rectified Linear Unit (ReLU) as the loss function and activation function respectively, initialize all the weights using the imagenet's pre-trained model, optimize the network by ADAM algorithm, set the batch size to be 50, and adjust the learning rate to be 10$^{-4}$ for the first $10^4$ epoch and to be 10$^{-6}$ for another $10^4$ epoch. The training process lasts several hours for Alexnet and twenty hours for VGG16 with GPU RTX 2080 Ti single card.

\section{Simulation and results} \label{Simu-res}
\subsection{The toy model}
We test two cases: one is the noise case ($n\ne 0$ with different noise levels) used to test the consistency between Eq.(\ref{eq_error1}) and Eq.(\ref {eq_error3}), and the no-noise ($n=0$) to test Eq. (\ref{eq_error2}) and Eq.(\ref {eq_error3}). For both cases, $d$ is drown from a uniform distribution U[-2, 2]. For the noise case, Gaussian noise $n\sim$N(0, $\sigma$) is adopted to generate a sample $X=d+n$  with three labels $Y_i$, which have the values $\{{Y_1=d}, {Y_2=d^2}, {Y_3=d^3}\}$. To avoid the over-fitting issue, 200,000 $X$ with $\sigma=0.05$ and 20,000 $X$ with $\sigma=0.5$ are adopted for training and testing processes, respectively. %\textcolor{red}{It is worth to nothing that, the feature distribution of the training set and the test set is not independent and identically distributed. The noise level of the training set is smaller than that of the test set. This is because we want to ensure that the model can learn accurate mappings (Eq. \ref{eq_dataxy})}. 
It is worth to note that, the feature distribution of the training set and the test set is not independent and identically distributed. The noise level of the training set is smaller than that of the test set. This is because we want to ensure that the model can learn accurate mappings (Eq.(\ref{eq_dataxy})). The function $\mathcal{M}$  is calculated by the standard deviation of predictions $\{\sigma_{(\hat{Y_1})},\sigma_{(\hat{Y_2})},\sigma_{(\hat{Y_3})}\}$, which are evaluated in 20 bins of $d$.

\begin{figure*} %[!htb]
	\centering
	\includegraphics[height=5cm,width=16.0cm]{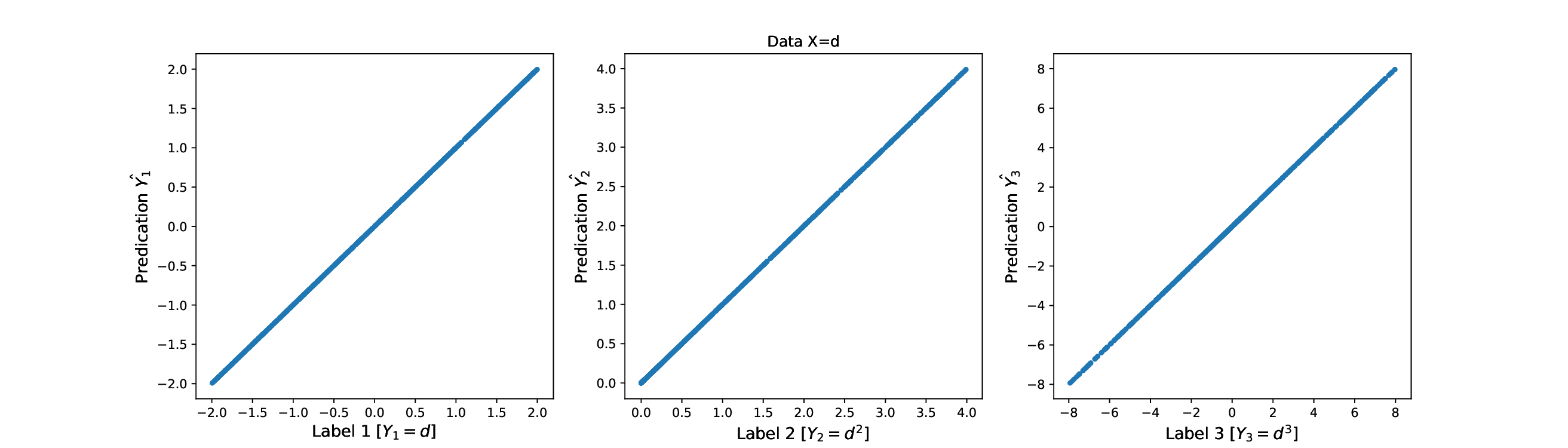}
		\includegraphics[height=5cm,width=16.0cm]{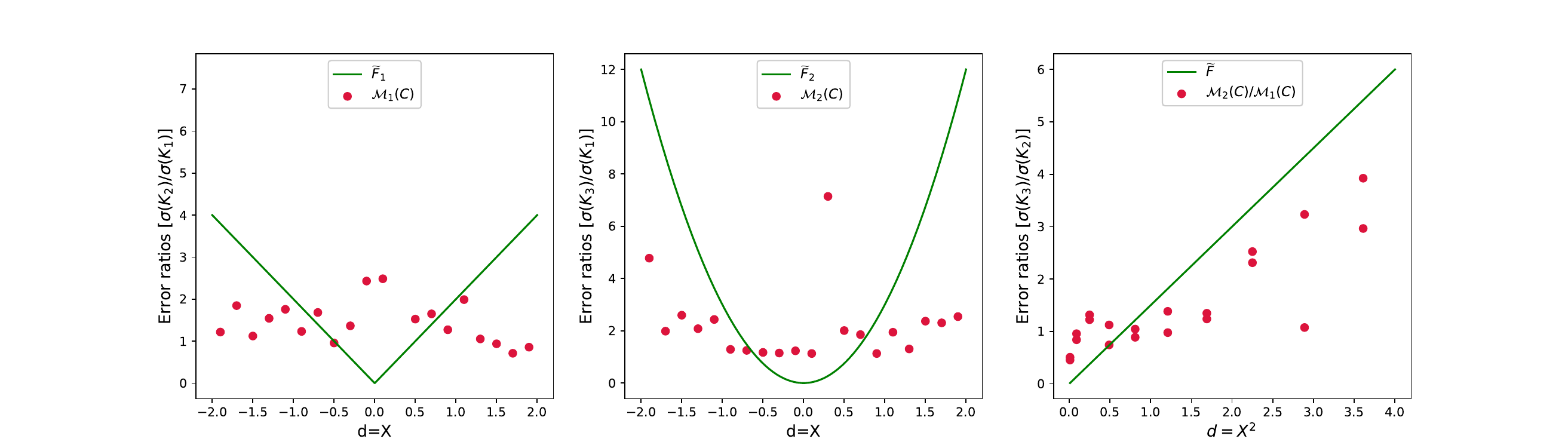}
			\includegraphics[height=5cm,width=16.0cm]{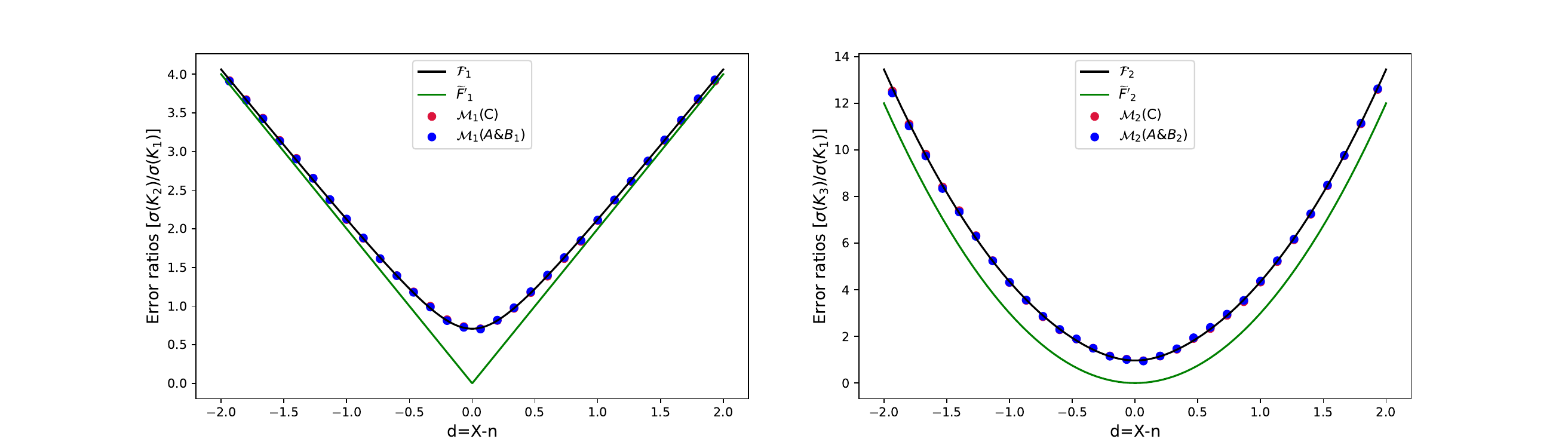}
	\caption{Predictions of the Network and comparisons of the function $\mathcal{M}$ (red points), $\mathcal{F}$ (blue lines) and $\widetilde{\mathcal{F}}$ (green lines). The function $\mathcal{M}$ are shown by the standard deviation of predictions $\{\sigma_{(\hat{Y_1})}, \sigma_{(\hat{Y_2})}, \sigma_{(\hat{Y_3})}\}$, which are evaluated in 20 bins of $d$. The up-panel and middle-panel are for no noise $n=0$ case, and down-panel is for noise case. For function $\mathcal{M}$ defined in Eq. (\ref{eq_error3}), three networks are adopted: $\mathcal{M}(C)$ is for $\{\sigma_{(\hat{Y_1})},\sigma_{(\hat{Y_2})},\sigma_{(\hat{Y_3})}\}$ by Network $C$; $\mathcal{M}$(A$\&$B1) is for $\{\sigma_{(\hat{Y_1})}\}$ by Network $A$ and $\{\sigma_{(\hat{Y_2})}\}$ by Network $B1$, while $\mathcal{M}$(A$\&$B2) is for $\{\sigma_{(\hat{Y_1})}\}$ by Network $A$ and $\{\sigma_{(\hat{Y_3})}\}$ by Network $B2$.}
	\label{tool}
\end{figure*}

Fig \ref{tool} shows the prediction results of the toy model for the no-noise case (in up-panel) and comparison of the function $\mathcal{M}$ (red points, Eq.(\ref {eq_error3})), $\mathcal{F}$ (blue lines, Eq.(\ref {eq_error1})) and $\widetilde{\mathcal{F}}$ (green lines, Eq.(\ref {eq_error2})) for both no-noise (in middle-panel) and noise cases (in down-panel). For the no noise ($n=0$) case, according to Eq.(\ref{eq_dataxy}) $X=Y$, model prediction errors $\{N_1, N_2, N_3\}$ are only determined by the network itself, and the predictions of the Network almost do hot have errors ($\{N_1, N_2, N_3\} \sim 0$ shown in the up-panel of Fig.\ref{tool}). The propagation of errors of $\hat{Y}$ caused by unknown network prediction mechanism (the function $\mathcal{M}$ in the middle-panel-panel of Fig.\ref{tool})  do not follow the law of error propagation, e.g., $\mathcal{M}_1 (\sigma_{(\hat{Y_2})}, \sigma_{(\hat{Y_1})})\ne\widetilde{\mathcal{F}}_1$. 
On the other hand, for the noise $n\ne0$ case shown in down-panel of Fig.\ref{tool}, the propagation of errors of $\hat{Y}$ follows the law of error propagation quite well, e.g. $\mathcal{M}_1 (\sigma_{(\hat{Y_2})}, \sigma_{(\hat{Y_1})})\sim\mathcal{F}_1$  and $\mathcal{M}_2 (\sigma_{(\hat{Y_3})}, \sigma_{(\hat{Y_1})})\sim\mathcal{F}_2$ for all noise level. Note that, since we fixed the noise level $\sigma$, larger $|d|$ represents smaller noise case (Eq.(\ref{eq_error2})). Those trends do not depend on the predication parameters set, since Network A, B and C return almost identical results.

The results of noise case $\mathcal{M} \sim\mathcal{F}$ indicate that the network ''knows'' the distribution of noise $n$ and the relation of label numbers $\{{Y_1},{Y_2}, {Y_3}\}$, although there is no information on $n$ in the labels and the relation itself is also not included in the labels. Although we do not know the  predication mechanism of the neural network for parameters predication but only the structure of the networks, it seems that the neural network could guarantee the association among parameters. This performance of the neural network could further  implies  as the neural network is capable to learn the relationship among parameters. Those properties of Network are more interesting when it is applying for more sophisticated physics models, e.g., the gravitational lens model.   

\subsection{the SIE lens model}\label{LensPre}
Following \cite{nature}, we consider the singular isothermal ellipse (SIE) lens model (Eq.\ref{m_e}) and fix the redshift of lens $z_{l}=0.5$, the redshift of source $z_{s}=2$ and adopt 737 cosmology model (i.e.,$h=0.7,\Omega_{m}=0.3, \Omega_{\Lambda}=0.7$) \citep{737com}. The values of Einstein's radius $\theta_{E}$, the complex ellipticity ($\epsilon_{x},\epsilon_{y}$) and the position of lens centre ($x,y$) for both training and testing data are drawn from the uniform distribution shown in Table \ref{Errors}.
We simulated the lensed images for training and testing based on source images from COSMOS$-$23.5, COSMOS$-$25.2 in GREAT3 data. To test the generalization of networks trained by the images with high quality from GREAT3, we also use data from Galaxy Zoo as source image to produce another test dataset. With the VGG16 and AlexNet trained by GREAT3 data (two million samples in total), we estimate the parameters \{$\theta_{E}$, $\epsilon_{x}$, $\epsilon_{y}$, $x$, $y$\} of other branches of GREAT3 data (ten thousands samples in total), labeled such as VGG16($\theta_{E}$) and Alexnet($\theta_{E}$) in Table~\ref{Errors}, respectively. More details on data, training, testing, robustness,accuracy of individual parameter estimation and error propagation can be found in the following content.

%The effective lens mass $ M_{L}$ within the Einstein's radius $\theta_{E}$ is calculated by
%\begin{equation}\label{m_e}
%M_{L}=\frac{c^{2}\theta_{E}^{2}D_{l}D_{s}}{4G D_{ls}}
%\end{equation}
%where $D_{l}, D_{s}$ are the angular diameter distance of lens and source respectively, $D_{ls}$ is the angular diameter distance between lens and source.

The source images for training data are from COSMOS$-$23.5 and COSMOS$-$25.2.
All source images are first convolved by the point spread function (PSF) supported in GREAT3 data to improve image quality. These images are used to produce two millions lensed images with parameters shown in Table \ref{Errors}. Each lensed image undergoes the following operations before being fed into the network to avoid overfitting. First, add random Gaussian noise to the lensed image. The root mean square value of the noise is randomly selected from a uniform distribution, and its value is 1\% -10\% of the signal. Then, we use a factor of 50–1,000 to convert the image to a photon count, and use these values as the $\lambda$ to generate a Poisson realization map, effectively adding Poisson noise to the image. We use the 400,000 images including simulated hot pixel and cosmic rays provided by \cite{nature} to make the network insensitive to pixel artifacts and cosmic rays. Then we use a random root mean square Gaussian filter to convolve the image to simulate the blurring effect of the PSF that reveals the factors of atmosphere and the telescope itself. Finally, randomly translate on the image for augmenting the data. The total training data samples are two million in total and in which ten thousand datasets are used for validation.   Each  training data sample is fed into the neural network with  different data augmentation operations.

\begin{figure*}
        \subfloat[]{
	\begin{minipage}[t]{0.495\linewidth}
	\includegraphics[height=5cm,width=8.0cm]{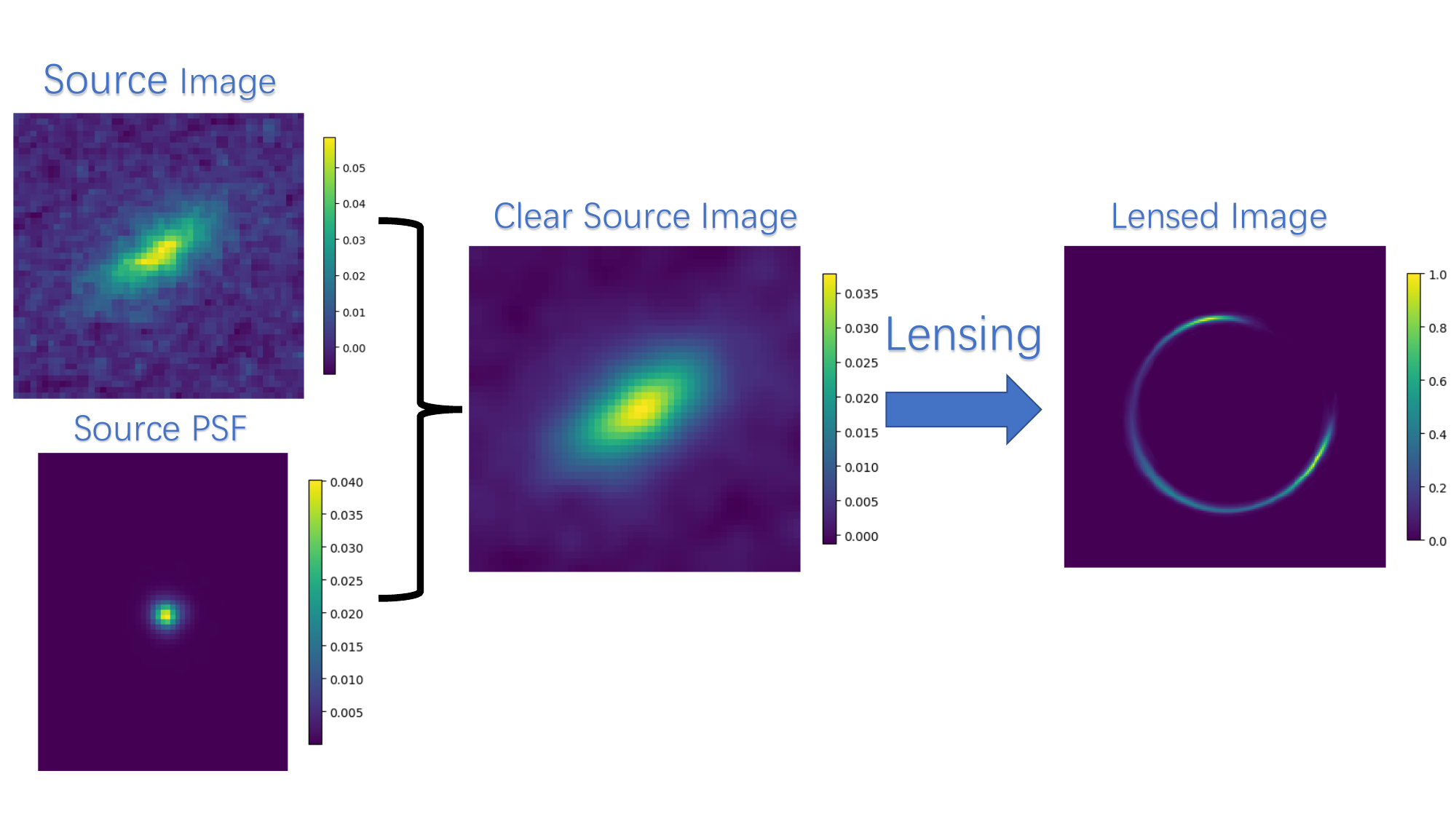}
        \end{minipage}
        }
        \subfloat[]{
	\begin{minipage}[t]{0.495\linewidth}
	\includegraphics[height=5cm,width=8.0cm]{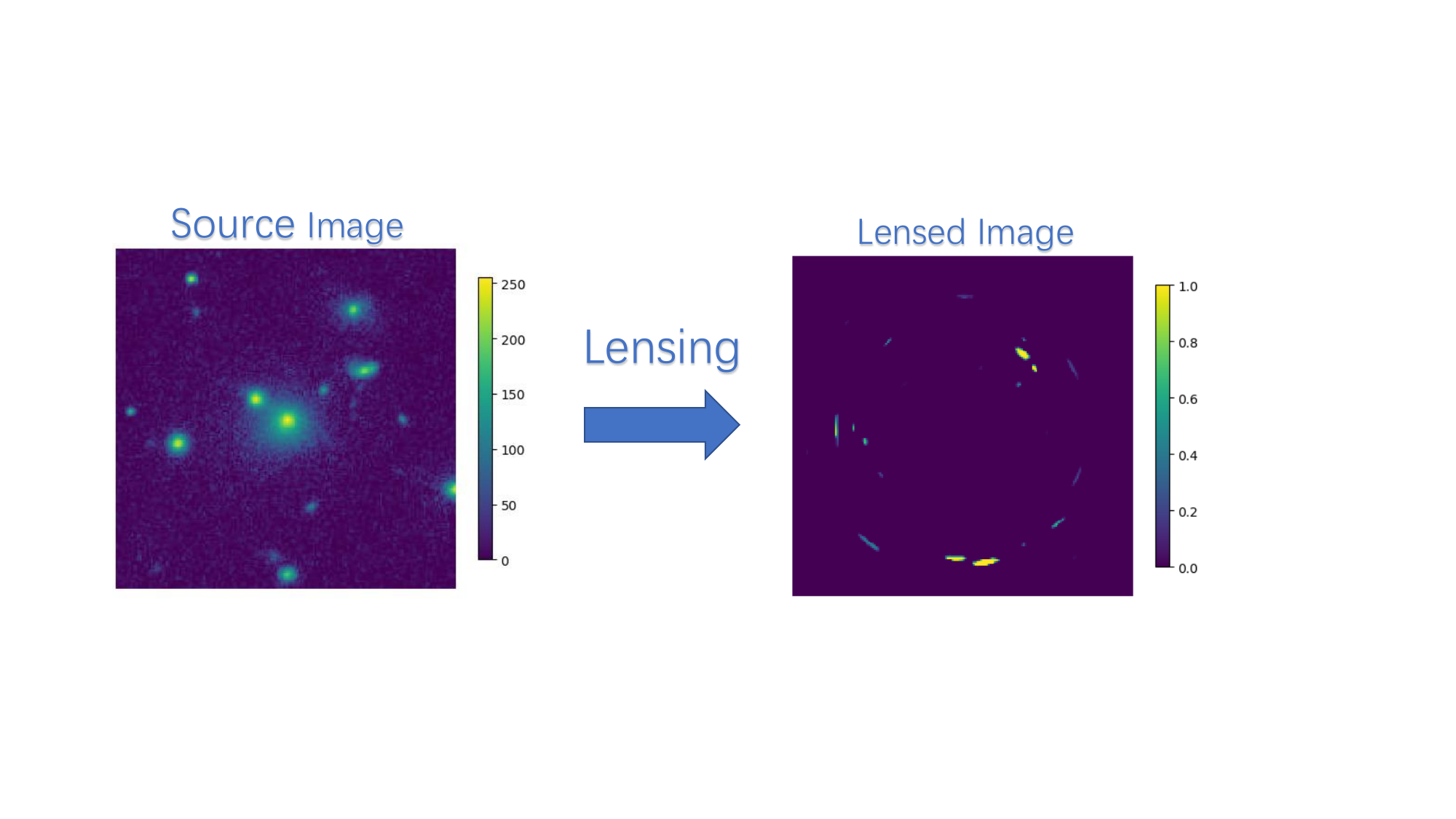}
        \end{minipage}
        }
        
	\caption{(a) One simulated training data-set in GREAT3 data group; (b) One simulated test dataset in Galaxy zoo data group.
	\label{figdata}}
\end{figure*}

We use other branch of GREAT3 data (1.8 million samples) as source images to produce our test dataset (ten thousands samples in total). To test the generalization of networks trained by the high quality of image from GREAT3, we use data from Galaxy Zoo (61 thousands samples) as source image to produce another test dataset (ten thousands samples in total). The data in the Galaxy Zoo are coming from the Sloan Digital Sky Survey (SDSS). There may be multiple galaxies and a higher noise level comparing to the GREAT3 data. One training image from GREAT3 and one test image from Galaxy Zoo are illuminated in Fig.\ref{figdata}(a) and Fig.\ref{figdata}(b), respectively.
%%%%%%%%%%%
%%%%%%%%%%%
To understand the robustness of networks, we use the VGG16($M_{L}$) and Alexnet($M_{L}$), which are trained by the GREAT3 datasets with higher image quality, to predict lens mass $M_{L}$ of test datasets in the Galaxy zoo data group.

Shown in Table \ref{Errors}, the standard deviations of the parameters estimated by VGG16 are all slightly better than Alexnet.
It should be noted that in \cite{nature} the errors of their Alexnet networks seem to be much better than ours. The reason is that the test datasets in \cite{nature} are unknown for us, we get corresponding results only considering the network (AlexNet) with their trained weights.  The standard deviation of \{$\epsilon_{x}$, $\epsilon_{y}$, $x$, $y$\} from VGG16($M_{L}$) are comparable to the results from VGG16($\theta_{E}$).  

\begin{figure*}
        \subfloat[]{
	\begin{minipage}[t]{0.495\linewidth}
	\includegraphics[height=5cm,width=8.0cm]{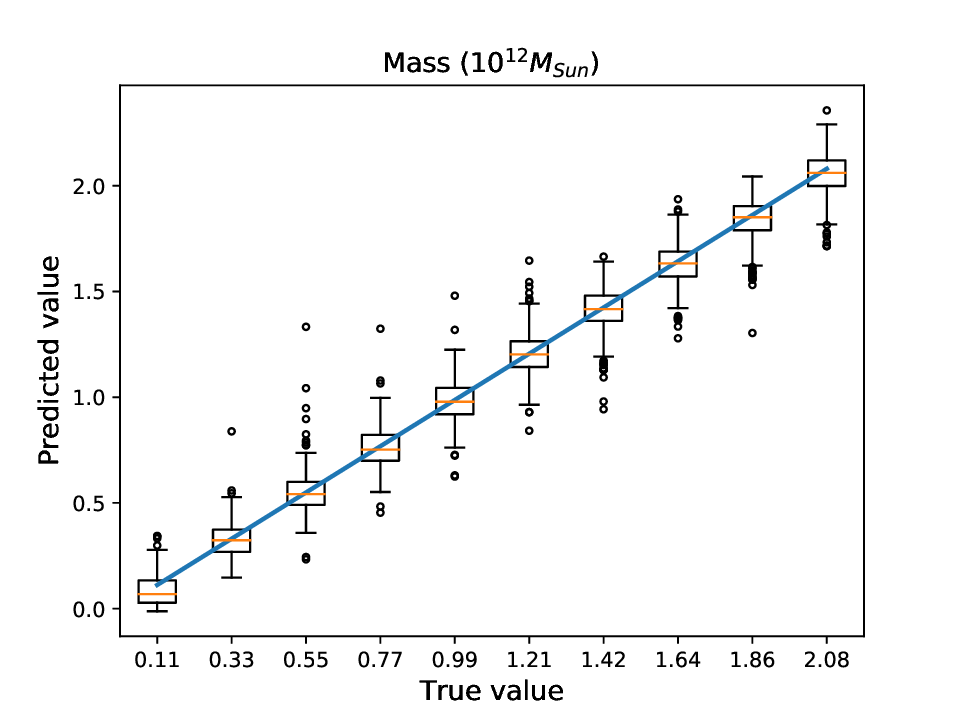}
        \end{minipage}
        }
        \subfloat[]{
	\begin{minipage}[t]{0.495\linewidth}
	\includegraphics[height=5cm,width=8.0cm]{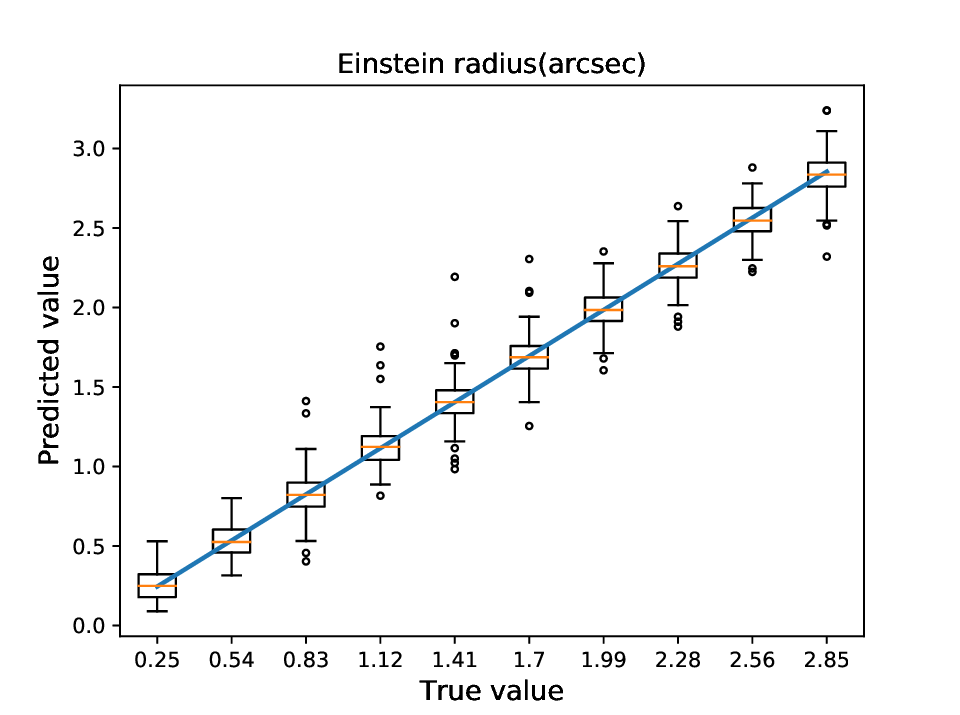}
	\end{minipage}
        }
        
	\caption{Results for the GREAT3 data. The estimated values are shown on the $y$ axis, while the $x$ axis represents the segment of the point. (a)Comparison of estimated lens masses with their true values by VGG16($M_{L}$) network. (b)Comparison of estimated lens Einstein radius with their true values by VGG16($\theta_{E}$) network. }
	\label{figMl}
\end{figure*}

\begin{figure}
	\centering
%	\subfigure[]{
%		\includegraphics[height=5cm,width=8cm]{VGG_Ml_galaxy.eps}}
%	\subfigure[]{
%		\includegraphics[height=5cm,width=8cm]{Galaxyzoo_VGG_res.eps}}
%	\subfigure[]{
%		\includegraphics[height=5cm,width=8.0cm]{VGG_find_why_small2.eps}}
%	\quad
%	\subfigure[]{
%		\includegraphics[height=5cm,width=8cm]{VGG_find_why_small1.eps}}
%	\subfigure[]{
%		\includegraphics[height=5cm,width=8.0cm]{VGG_find_why_big2.eps}}
%	\subfigure[]{
%		\includegraphics[height=5cm,width=8cm]{VGG_find_why_big1.eps}}\\

        \subfloat[]{
	\begin{minipage}[t]{0.495\linewidth}
		\includegraphics[height=5cm,width=8cm]{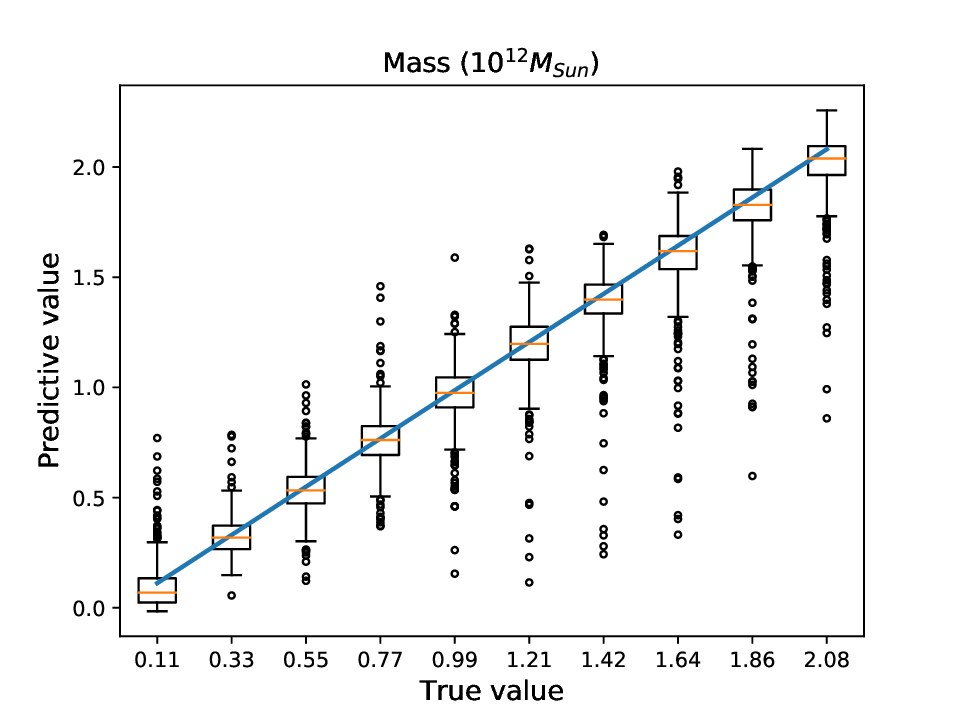}
        \end{minipage}
        }
        \subfloat[]{
        \begin{minipage}[t]{0.495\linewidth}
		\includegraphics[height=5cm,width=8cm]{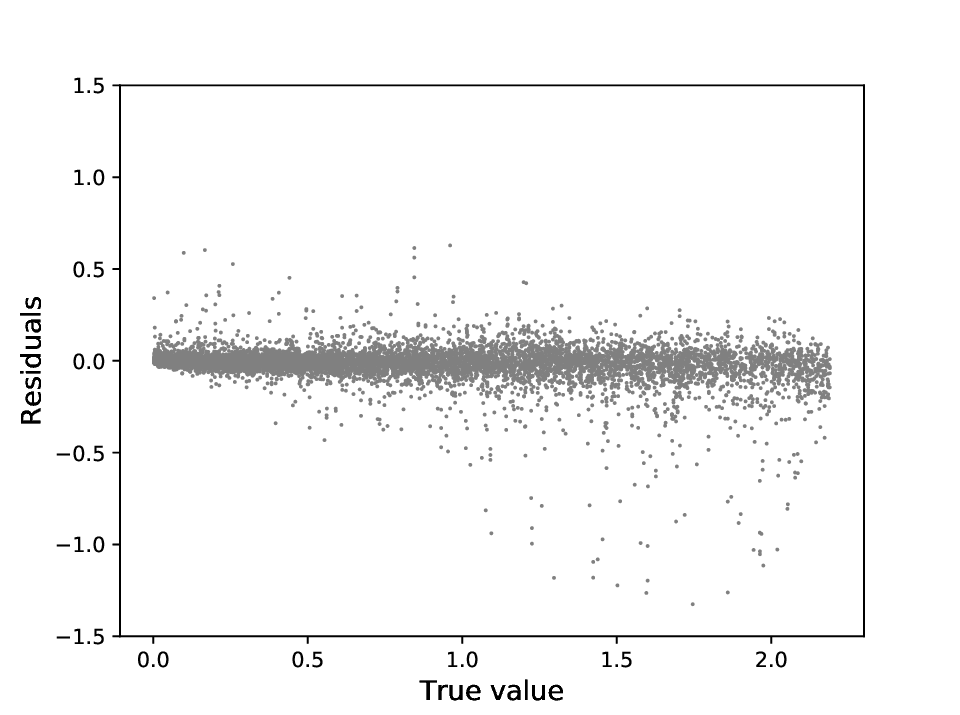}
	\end{minipage}
        }
        
	\subfloat[]{
        \begin{minipage}[t]{0.495\linewidth}
		\includegraphics[height=5cm,width=8cm]{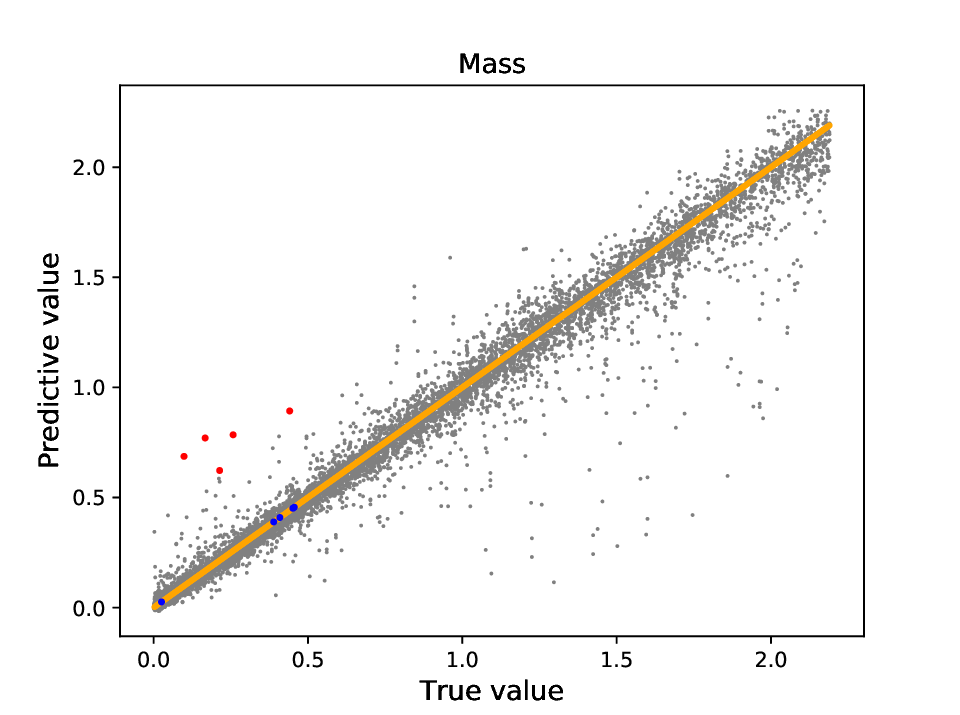}
        \end{minipage}
        }
        \subfloat[]{
        \begin{minipage}[t]{0.495\linewidth}
		\includegraphics[height=5cm,width=8cm]{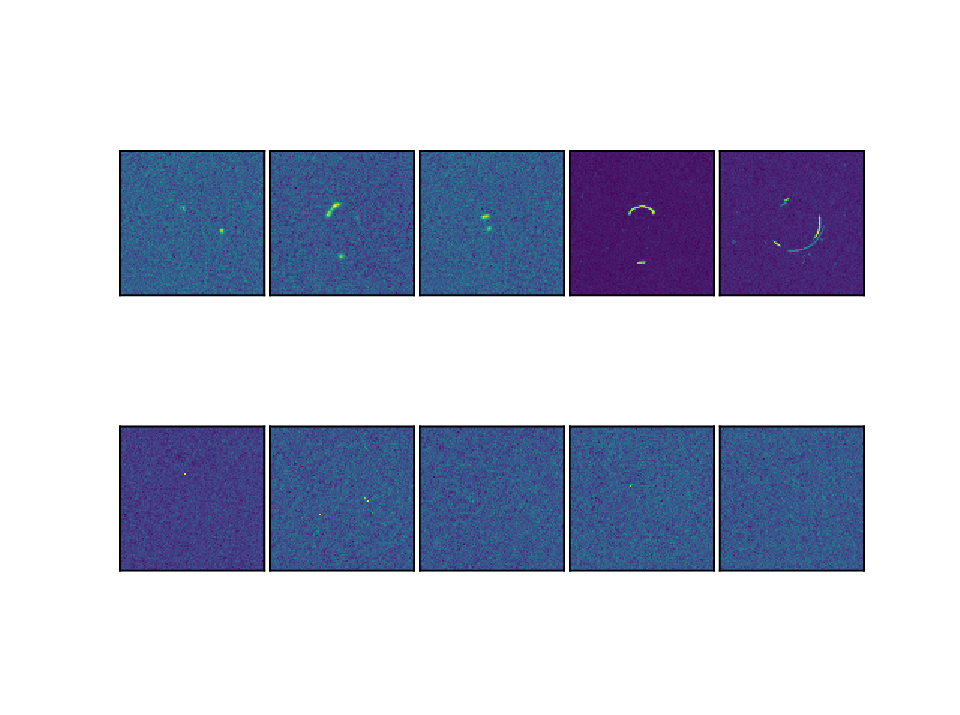}
        \end{minipage}
        }
        
        \subfloat[]{
        \begin{minipage}[t]{0.495\linewidth}
		\includegraphics[height=5cm,width=8cm]{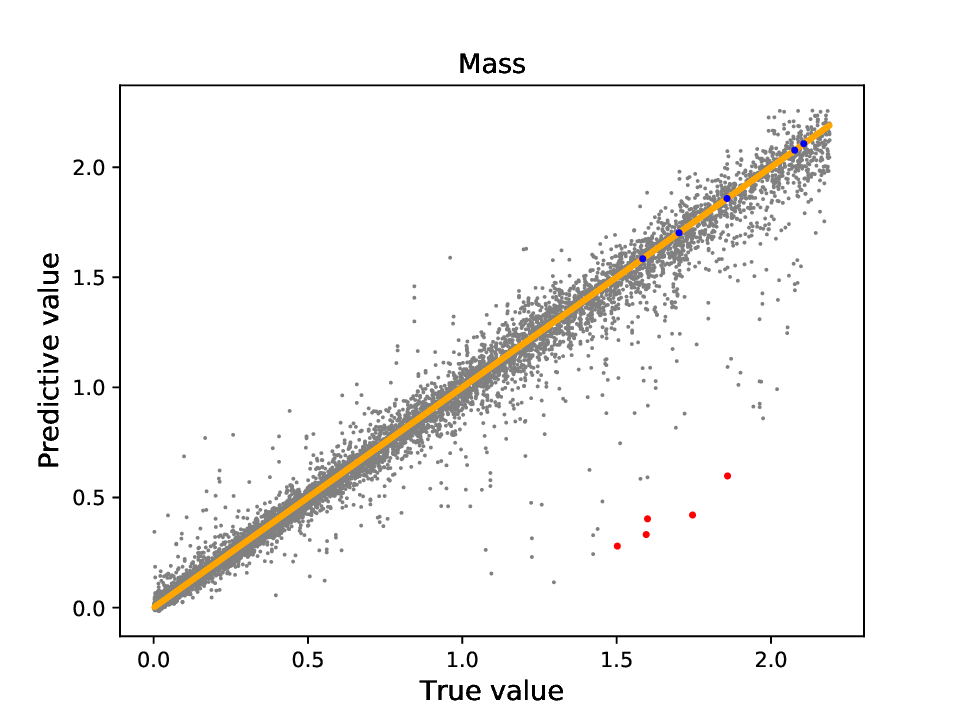}
        \end{minipage}
        }
        \subfloat[]{
        \begin{minipage}[t]{0.495\linewidth}
		\includegraphics[height=5cm,width=8cm]{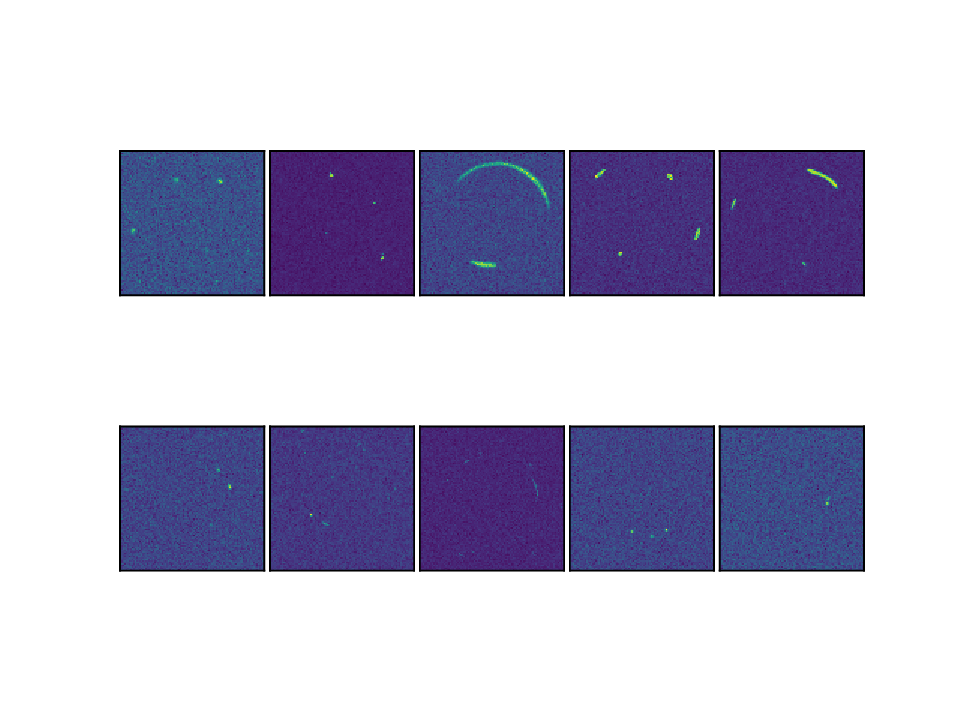}
        \end{minipage}
        }
        	\caption{Results for the Galaxy zoo data by VGG16($M_{L}$). The reason for the predicted value of large mass being less than the real value. (a) The box plot of the estimated lens masses compared to the true value. (b) The residuals plot of lens Mass prediction. (c) The comparison of estimated lens masses with their true values, in which the red dots are the poor estimation samples corresponding the images in the second row of (d), the blue dots are perfectly estimated samples corresponding the images in the first row of (d). (e) The comparison of estimated lens masses with their true values, in which the red dots are the poor estimation samples corresponding the images in the second row of (f), the blue dots are perfectly estimated samples corresponding the images in the first row of (f).}
	\label{figgen2}
\end{figure}

To check if the predication is also depended on the parameter value, we compare the estimated lens masses $\hat{M_L}$ by VGG16($M_{L}$) and Einstein's radius $\hat{\theta_{E}}$ by VGG16($\theta_{E}$) with their true values with box plot for the VGG16 trained by GREAT3 data (Fig. \ref{figMl}). We divide the interval into 10 segments with equal width, and draw its box plot for each segment. For the box plot of the Einstein's radius, every bin has the same data approximately. But for the box plot of the lens mass, there are more data in the bins with small mass, because the mass is proportional to the square of the Einstein's radius. Shown in Fig.\ref{figMl}, the mean value of predicted mass $\hat{M_L}$ by VGG16($M_{L}$) recover better the true value $M_L$, although there are more outliers than the Einstein's radius $\hat{\theta_{E}}$ by VGG16($\theta_{E}$). For massive galaxies more outliers are in smaller prediction value comparing with the true value, while for less massive galaxies more outliers are in lager predicted value. 
The results from the Galaxy zoo data are shown in Fig.\ref{figgen2}. 
 The value of $M_{L}$ and the residuals of lens mass predicted by VGG16($M_{L}$) are shown in Fig.\ref{figgen2} (a) and (b), respectively. Although the average value of predicated parameters represent the true value quite well, there are more outliers resulting larger standard deviations for both networks (see Table \ref{Errors}). 
This is partly because the datasets in the Galaxy zoo data group are sources with irregular shapes and have very noisy background as the image shown in Fig.\ref{figdata}(b).
It can be seen that the reason for clustering to the average value is that the noise of the image is too large to contain useful information (see the image data of outliers in Fig.\ref{figgen2} (d, f) corresponding to the outliers in Fig.\ref{figgen2} (c, e)). 
In order to minimize the overall loss, neural networks tend to output the average value of the sample. The detailed comparison of estimated lens masses with their true values is figured out in Fig.\ref{figgen2}, which indicates the predicted value of small mass tends to be greater than the real value, while the predicted value of large mass is less than the real value (also see the residuals plot in Fig. \ref{figgen2} (b)). This tendency is also found for \{$\epsilon_{x}$, $\epsilon_{y}$, $x$, $y$\}.

\begin{figure}
	\centering
	%\subfigure[]{
	%	\includegraphics[height=5cm,width=8.0cm]{error_pap.eps}}
	%\subfigure[]{
	%	\includegraphics[height=5cm,width=8.0cm]{error_pap2.eps}}\\
%		\gridline{\fig{error_pap_1.eps}{0.5\textwidth}{(a: by  VGG16($\theta_{E},M_{L}$))}
%		\fig{error_pap_2.eps}{0.5\textwidth}{(b: by  VGG16($\theta_{E},M_{L}$)))}
%	}
%	\gridline{\fig{error_pap_3.eps}{0.5\textwidth}{(c: by VGG16($\theta_{E}$)  and VGG16($M_{L}$) )}
%		\fig{error_pap_4.eps}{0.5\textwidth}{(d: by VGG16($\theta_{E}$)  and VGG16($M_{L}$) )}
%	}
    \subfloat[]{
	\begin{minipage}[t]{0.495\linewidth}
		\includegraphics[height=5cm,width=7.5cm]{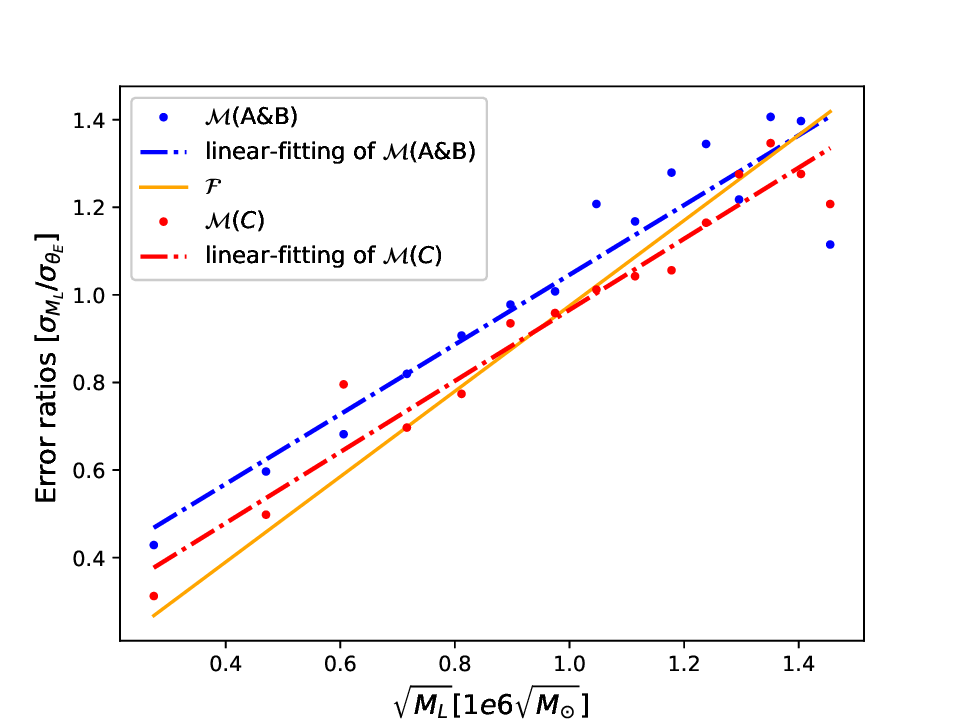}
        \end{minipage}
        }
        \subfloat[]{
	\begin{minipage}[t]{0.495\linewidth}
		\includegraphics[height=5cm,width=7.5cm]{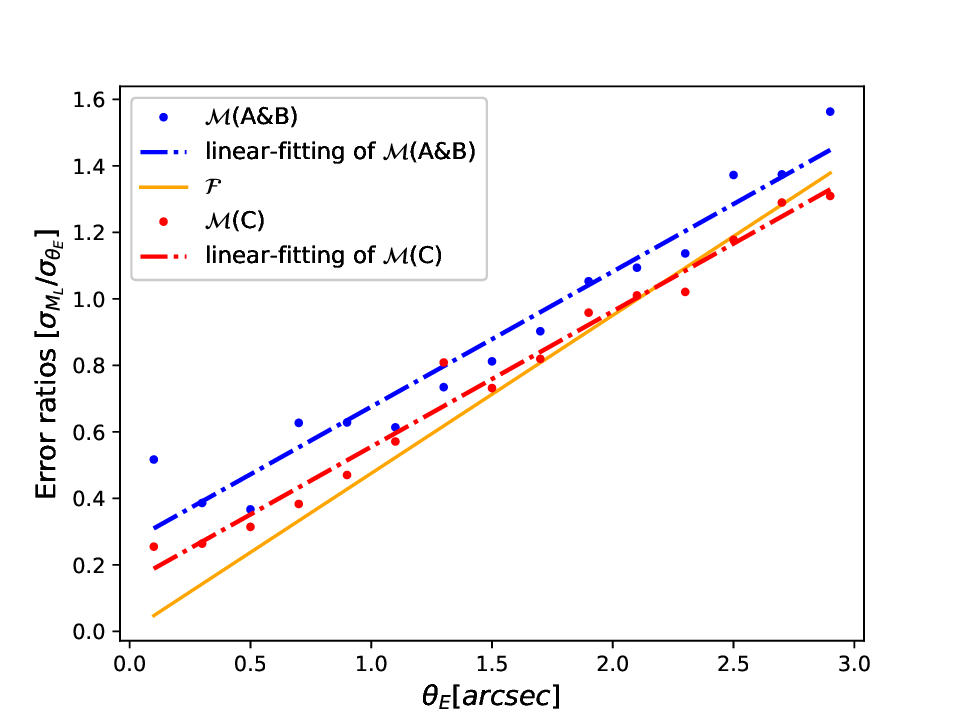}
	\end{minipage}
        }
	\caption{(a) The comparison of theoretical and predicted correlation between $\sqrt{M_{L}}$ and $\frac{\sigma_{\hat{M_{L}}}}{\sigma_{\hat{\theta_{E}}}}$. (b) The comparison of theoretical and predicted correlation between $\theta_{E}$ and $\frac{\sigma_{\hat{M_{L}}}}{\sigma_{\hat{\theta_{E}}}}$. The blue dots ($\mathcal{M}$(A$\&$B)) represent the predicted values of VGG16($\theta_{E}$) and VGG16($M_{L}$), the yellow line ($\mathcal{F}$) represents the result from the theoretical error propagation formula (Eq. (\ref{err_p})) and the red line (linear-fitting of $\mathcal{M}$(C)) represents the linear-fitting of the red dots from the predicted values of VGG16 ($\theta_{E}, M_{L}$).} 
	\label{figerror}
\end{figure}

The standard deviation of all parameters predicated by VGG16 or Alexnet in three sets of labels for all test samples are shown in Table~\ref{Errors}. In order to investigate the relation among the prediction errors of $\theta_{E}$ and $ M_{L}$, 
the ratio of the standard deviation of $\hat{M_L}$ and $\hat{\theta_E}$ by VGG16 networks A, B, C (test dataset in 15 segments) as a function of the center value of $M_{L}$ is shown in Fig. \ref{figerror}. 
 The results show that 
\begin{equation}\label{err_p2}
\mathcal{M}\equiv \frac{\sigma_{\hat{M_{L}^{C}}}}{\sigma_{\hat{\theta_{E}^{C}}}}  \sim c\sqrt{\frac{D}{G}}\sqrt{M_{L}} \ne    \frac{\sigma_{\hat{M_{L}^{B}}}}{\sigma_{\hat{\theta_{E}^{A}}}} .
\end{equation}
It can be found that the error propagation by VGG16($\theta_{E},M_{L}$) is roughly consistent with the error propagation formula (the yellow line represent Eq.~(\ref{err_p})  $\mathcal{F}$ in Fig. \ref{figerror}): $\mathcal{M} \sim\mathcal{F}$ as shown in toy model case. Again, the network C seems "know" the noise distribution since the errors follows the theoretical error propagation formula. However, the ratio of the errors of $\hat{\theta_{E}}$ and $\hat{M_{L}}$ derived from the network A and B does not follow the theoretical error propagation formula, which is not the case in toy model. The plausible reason for the difference between toy model and lens model is that Eq.~(\ref{err_p}) does not consider the non-Gaussian noise effects of input data.

This result enlightens us that as long as the accuracy of parameter estimation by the network is guaranteed, even if we do not know the physical relation between the parameters of input data, the relation will be reflected through the corresponding error of deep learning estimation. This feature of deep learning  is valuable for further investigation on the parameter correlation with unknown theoretical model in advance.  For example, if the lens mass is measured by gravitational wave \cite{2020PhRvD.101f4011H}, one could combine the lens mass estimation from gravitational wave, Einstein radius estimation  from optical lens image  and the redshifts $z$ from emission lines to investigate  $M_L - \theta_{E}-z$ relation.
%These results inspire us to investigate the correlation of each parameter in the model using machine learning.

\begin{table}[h]
	\centering
	\caption{The standard deviation of the parameter predictions for the test datasets in GREAT3 and galaxy zoo. The angular parameters ($\theta_{E}$, $x$ and $y$) are given in units of arcseconds, lens mass $M_{L}$ is in unit of $10^{12}M_{\odot}$. The results of the Alexnet in \cite{nature} are obtained by applying their trained weights to test datasets. VGG16($\theta_{E}$) and Alexnet($\theta_{E}$) are trained to predict the \{$\theta_{E}$, $\epsilon_{x}$, $\epsilon_{y}$, $x$, $y$\}. The VGG16($M_{L}$) and Alexnet($M_{L}$) are for \{$M_{L}$, $\epsilon_{x}$, $\epsilon_{y}$, $x$, $y$\}, and VGG16($\theta_E, M_{L}$) is used for \{$\theta_{E}$, $M_{L}$, $\epsilon_{x}$, $\epsilon_{y}$, $x$, $y$\}. The ranges of uniform distribution of lens model parameters for networks are shown in square brackets under the parameters. The value of mass $M_{L}$ is calculated by Eq.~(\ref{m_e}) given $\theta_{E}$.}
	\label{Errors}
	\begin{tabular}{l  lllllll}\\
		\hline
	Test datasets&	Network & $\theta_{E}$~~~& $M_{L}$~~ & ~~$\epsilon_{x}$& ~~$\epsilon_{y}$ &~~$x$ &~~$y$ \\
		&	 &[0,~3.0]~~&[0,~2.19] &[0,~0.9]&  [0, 0.9] &[-0.25,~0.25]  &[-0.25,~0.25]  \\
		\hline
		GREAT3&		 VGG16 ($\theta_{E}$) &0.047~ & ------- & ~0.080~ & ~0.071~ & ~0.075~& ~0.073\\
		&  Alexnet ($\theta_{E}$) &0.067~ & ------- & ~0.081~ & ~0.081~ & ~0.093~& ~0.091\\
		&VGG16 ($M_{L}$) &------- &0.048~ & ~0.095~ & ~0.086~ & ~0.089~& ~0.086\\
		&VGG16 ($\theta_{E}, M_{L}$) &0.050 &0.046~ & ~0.090~ & ~0.082~ & ~0.078~& ~0.079\\
		\hline
		\hline
		galaxy zoo&	VGG16($M_{L}$)&------- &0.097~& ~0.149~ & ~0.143~ & ~0.114~& ~0.111\\
		&	Alexnet($M_{L}$) &-------&0.141~& ~0.143~ & ~0.132~ & ~0.144~& ~0.143\\
		\hline
	\end{tabular}
\end{table}

\section{DISCUSSION AND SUMMARY}\label{summary}
Unlike the traditional parameter estimation, the parameter estimation by machine learning almost completely depends on the information in samples. Assuming the SIE lens model, the Einstein radius $\theta_{E}$ and effective lens mass $M_{L}$ are estimated by convolutional neural network, and the capability of network acquiring the correlation information between parameters from the data is tested through the estimation errors. In this process, 
the Networks produce the relation of errors as the traditional error propagation law based on known $\theta_{E}$-$M_{L}$ relation. Such correlation of estimated parameters provides a self-consistent result, so it is very important for further study on parameter estimation by machine learning. 

In order to ensure the reliability of the above results, the accuracy of parameter estimation by the network also needs to be guaranteed.  The convolutional neural network AlexNet is an effective approach of predicating  parameters of lens model (\cite{nature}). Through applying the typical convolutional neural network VGG on the parameter estimation of the gravitational lens, the great performance on abstraction of features has been shown in our simulated lens data (results are shown in Table \ref{Errors}  and  Fig \ref{figMl} and \ref{figgen2}). Meanwhile, the robustness of such a network could also be guaranteed to a certain extent. From the results of Galaxy Zoo test datasets, it is found that for the signal submerged in noise, neural networks tend to output the average of the training set to minimize the mean squared error (MSE). 
We also test the non-normal loss (MAE) as the loss function after the normal generative process, and the performance of the error propagation is similar as shown in the MSE case. Further study could also test more advanced networks on the performance of parameter estimation, such as ResNet (\cite{resnet}), DenseNet (\cite{densenet}), ViT (\cite{vit}), so as to get a more accurate error propagation corelation among parameters.

Although the MSE and MAE loss functions seem to only guarantee the accuracy of each parameter, they ensure that the model fits the functional relationship  between input data and output data. According to Universal Approximation theorem (\cite{1989Multilayer}), the functional relationship should be presented by the perfect network structure and weights of neurons.
%Based on a toy model, we elaborate the reasons for the neural network being not only able to learn the relation between samples and parameters, but also to learn the error propagation relation among parameters.  This is meaningful for the work about efficient inverting of physical information. (考虑是否写这句话)
The fact that the error of each parameter's estimation by  machine learning satisfies the error propagation formula is worth discussing. The general plausible reason for  this  consistency is  that the mapping exists between the predicted parameters and the input data. If there is noise in the input, the model will output the biased prediction containing noise according to the accurate mapping. Therefore, different parameters will exhibit the law of error propagation due to the same input noise. In particular, for the toy model, there is a simple functional relationship between input and output data i.e., $\{{\hat{Y_1}=X},{\hat{Y_2}=X^2}, {\hat{Y_3}=X^3}\}$. In training, the model trends to minimize the loss function in order to learn the functional relationship.  After training, the functional relationship is stored in the weight of each neuron and the entire model has the functional relationship contained in the training data. When we add the noise directly to $X$, since the new model's mapping is almost the same with the mapping contained in the training data, the predicted results exhibit the error propagation relationship. In lens model, it can be considered that the CNN layer performs feature extraction on the image. The output results of the last CNN layer are the latent variables representing the image. The fully connected layer is a function from latent variables to predicted results. The error of the image will cause the error of the latent variables, and the predicted values are function of the latent variables. So the error propagation formula is satisfied between the predicted values and the latent variables, and the error propagation formula is therefore satisfied between the predicted values.

\begin{acknowledgements}
We thanks Kai Liao for helpful discussions.
This work was supported by the National Natural Science Foundation of China (Grant Nos.11922303, 11873001), the Natural Science Foundation of Chongqing (Grant No. cstc2018jcyjAX0767), the Key Research Program of Xingtai 2020ZC005, and the Fundamental Research Funds for the Central Universities (grant No. 2042022kf1182).
\end{acknowledgements}

\bibliographystyle{raa}
\bibliography{ref}

\label{lastpage}

\end{document}